\documentclass[epj,twocolumn]{webofc}
\usepackage[varg]{txfonts}

\woctitle{ISVHECRI 2018}

\usepackage[displaymath, mathlines]{lineno}
\usepackage{xcolor}

\makeatletter
\newcounter{pagecount}
\newcommand{\limitpages}[1]{
    \setcounter{pagecount}{0}
    \gdef\maxpages{#1}
    \ifx\latex@outputpage\@undefined\relax
        \global\let\latex@outputpage\@outputpage
    \fi
    \gdef\@outputpage{
        \addtocounter{pagecount}{1}
        \ifnum\value{pagecount}>\maxpages\relax
        \else
            \latex@outputpage
        \fi
    }
}
\makeatother

\limitpages{4}

\begin{document}

\title{Atmospheric Muons Measured with IceCube}

\author{Dennis Soldin\inst{1}\fnsep\thanks{\email{soldin@udel.edu}} for the IceCube Collaboration\thanks{\protect\url{http://icecube.wisc.edu/collaboration/authors/current}}}

\institute{Bartol Research Institute and Dept. of Physics and Astronomy, University of Delaware, Newark, DE 19716, USA}

\abstract{IceCube is a cubic-kilometer Cherenkov detector in the deep ice at the geographic South Pole. The dominant event yield is produced by penetrating atmospheric muons with energies above several $100\,\mathrm{GeV}$. Due to its large detector volume, IceCube provides unique opportunities to study atmospheric muons with large statistics in detail. Measurements of the energy spectrum and the lateral separation distribution of muons offer insights into hadronic interactions during the air shower development and can be used to test hadronic models. 

We will present an overview of various measurements of atmospheric muons in IceCube, including the energy spectrum of muons between $10\,\mathrm{TeV}$ and $1\,\mathrm{PeV}$. This is used to derive an estimate of the prompt contribution of muons, originating from the decay of heavy (mainly charmed) hadrons and unflavored mesons. We will also present measurements of the lateral separation distributions of TeV muons between $150\,\mathrm{m}$ and $450\,\mathrm{m}$ for several initial cosmic ray energies between $1\,\mathrm{PeV}$ and $16\,\mathrm{PeV}$. Finally, the angular distribution of atmospheric muons in IceCube will be discussed.}

\maketitle

\section{Introduction}
\label{Sec:1}

IceCube is a cubic-kilometer neutrino detector installed in the ice at the geographic South Pole between depths of $1450\,\mathrm{m}$ and $2450\,\mathrm{m}$ \cite{IceCube}. Atmospheric muons produced in cosmic ray air showers above a muon energy threshold of roughly $460\,\mathrm{GeV}$, depending on their arrival directions, penetrate the Antarctic ice and trigger the deep ice detector with an average rate of about $2.15\,\mathrm{kHz}$. Taking data in its final detector configuration since 2011, IceCube therefore provides a unique dataset of high-energy atmospheric muons with enormous statistics. With its large $3$-dimensional detector volume IceCube observes muons from all directions and enables detailed studies of their angular distribution, up to zenith angles of approximately $82^\circ$. Thus, IceCube covers a very large phase space for the measurement of atmospheric muons, which is not accessible by any other existing experiment.

In this work we will present measurements of the energy spectrum of atmospheric muons above $10\,\mathrm{TeV}$. These measurements include detailed studies of the contribution of prompt muons, which originate from the decay of short-lived heavy hadrons. We will also present the lateral separation distribution of isolated muons far from the shower core, with separations up to several hundred meters. These muons are typically produced by decays of hadrons with large transverse momentum ($p_\mathrm{T}\gtrsim 2\,\mathrm{GeV/c}$) and thereby they provide tests of pQCD predictions and hadronic models at high energies and low Bjorken-x. Finally, the corresponding angular distribution of atmospheric muons measured in these analyses will be discussed.

In addition to the measurements presented in this work, IceCube's surface detector component IceTop \cite{IceTop} provides measurements of GeV muons at the surface, which are beyond the scope of this work and can be found in Ref. \cite{IceCube_RhoMu}. Using the deep ice detector together with the IceTop array also enables studies on forward muons with large Feynman-x, which can be found in Ref. \cite{IceCube_xF}, and measurements of the cosmic ray mass composition based on the ratio of the electromagnetic to muonic component of the air shower, which are presented in Ref. \cite{IceCube_comp}.

\section{High-energy muon fluxes}
\label{Sec:2}

High-energy (HE) muons are produced early during the development of cosmic ray air showers, mainly from the decay of pions and kaons. However, at high energies, above $\sim1\,\mathrm{PeV}$, the prompt contribution from leptonic decays of short-lived heavy hadrons and unflavored vector mesons is expected to dominate the total flux of atmospheric muons \cite{ERS,MCEq}. In IceCube high-energy muons are generally accompanied by a bundle of low-energy muons above threshold ($\gtrsim 460\,\mathrm{GeV}$), which forms the most compact region of the shower core. As shown in Ref. \cite{IceCube_HE1}, any bundle muon with an energy above $\sim10\,\mathrm{TeV}$ will presumably be the most energetic muon in the bundle.

The selection and the energy reconstruction of this most energetic muon is generally based on the energy loss characteristics in the ice. While at low energies the muon energy loss in the ice is highly dominated by continuous ionization, the contribution of stochastic (radiative) energy losses dominates towards higher energies. As described in Ref. \cite{IceCube_HE1}, the energy loss profile can therefore be used to select the most energetic muon and to estimate its energy at the surface in order to derive the HE muon spectrum. This has been done using two independent approaches: a cut-based event selection using two years of IceCube data \cite{IceCube_HE1} and a machine learning approach based on one year of data \cite{IceCube_HE2}.
\begin{figure}[tb]
\vspace{-0.1cm}

\mbox{\hspace{-0.5cm}\includegraphics[width=0.513\textwidth]{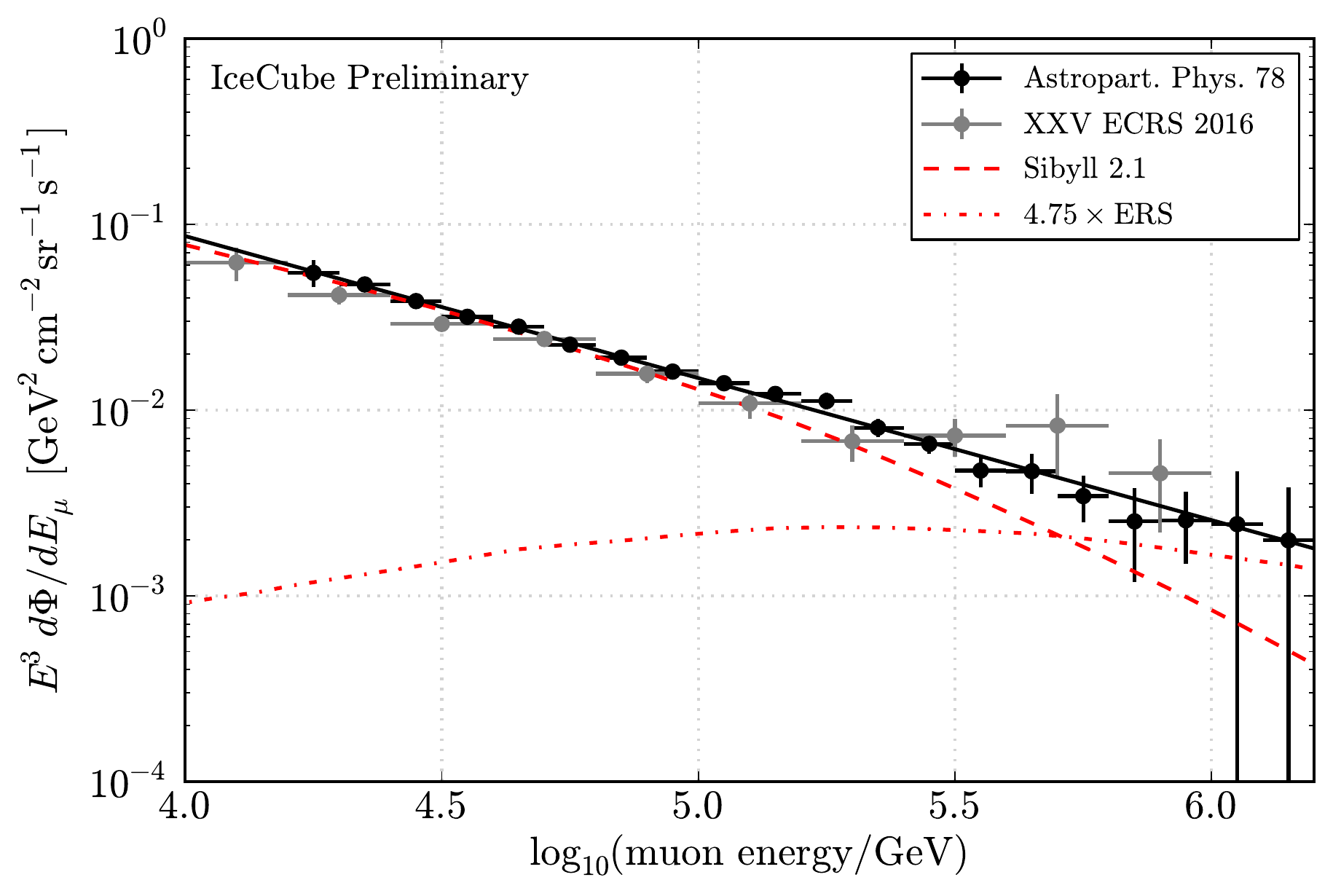}}
\vspace{-0.8cm}
\caption{Muon energy spectra measured in IceCube, taken from Ref. \cite{IceCube_HE1} (cut-based approach) and Ref. \cite{IceCube_HE2} (machine learning approach). The solid line represents the power law fit from Equation (\ref{eq:HE_spec}). Also shown are predictions of the conventional muon flux, obtained from CORSIKA using Sibyll 2.1 as hadronic model and assuming a H3a primary flux, as well as the best fit ERS prompt flux \cite{ERS} (see text for details).}
\label{fig:HE_spectrum}
\vspace{-0.8 cm}
\end{figure}
Figure \ref{fig:HE_spectrum} shows the resulting muon energy spectra at surface level ($\sim 690\,\mathrm{g/cm^2}$). Within the accuracy of these analyses, the spectrum can be  approximated by a simple power law of the form
\begin{linenomath}
\begin{align}
\frac{d\Phi}{dE_\mu}= \frac{(0.86 \pm 0.03) \cdot 10^{-10}}{\mathrm{TeV\,cm^{2}\,sr\,s}} \cdot \left( \frac{E_\mu}{10\,\mathrm{TeV}} \right)^{-3.76\pm0.02} ,
\label{eq:HE_spec}
\end{align}
\end{linenomath}
with $\chi^2/\mathrm{ndof}=5.36/18$ (solid line). Simulated Monte Carlo predictions obtained from the CORSIKA package \cite{CORSIKA} are also shown in Figure \ref{fig:HE_spectrum}, using Sibyll 2.1 \cite{Sibyll21} as hadronic interaction model and the H3a cosmic ray flux assumption from Ref. \cite{H3a}. CORSIKA predictions underestimate the experimental data towards high energies, which is expected to be caused by a missing prompt muon component in Sibyll 2.1. As described in Ref. \cite{IceCube_HE1}, this missing component is fit with multiples of the prompt ERS flux $\Phi_\mathrm{ERS}$ \cite{ERS}. Assuming an H3a primary flux, the best fit yields a prompt flux of $\Phi_\mathrm{prompt}=4.75\times \Phi_\mathrm{ERS}$. This estimate strongly depends on the underlying primary flux and the corresponding systematic uncertainties are therefore very large, ranging from $\Phi_\mathrm{prompt}=0.94\times \Phi_\mathrm{ERS}$ to $\Phi_\mathrm{prompt}=6.97\times \Phi_\mathrm{ERS}$. However, the non-existence of a prompt muon flux can be excluded with a significance of $1.52\sigma$ to $5.24\sigma$, depending on the primary flux assumption (see Ref. \cite{IceCube_HE1} for further details).

An improved analysis of the atmospheric HE muon spectrum, using three years of IceCube data, is in preparation. This analysis uses the most recent Sibyll 2.3 interaction model \cite{Sibyll23}, which includes a dedicated modeling of the prompt muon component. This will enable more sophisticated studies of the prompt contribution in simulations and, together with improved analysis methods and larger statistics, it will significantly reduce the uncertainties of the prompt muon flux estimate.

\section{Laterally separated muons}
\label{Sec:3}

In high-energy cosmic ray air showers hadrons with large transverse momentum $p_\mathrm{T}\gtrsim 2\,\mathrm{GeV/c}$ are produced which can subsequently decay into muons. These muons separate from the shower core while traveling to the ground, forming laterally separated (LS) muons with distances up to several $100\,\mathrm{m}$ from the dense core region. The resulting lateral separation is a direct measure of the $p_\mathrm{T}$ of the parent hadron. Experimentally a transition from soft to hard interactions is observed in the $p_\mathrm{T}$ spectrum, which falls off exponentially with a transition to a power law at approximately $2\,\mathrm{GeV/c}$, where interactions can be described in the context of pQCD \cite{Hagedorn}. This transition should be also visible in the lateral separation distribution of muons. 

\begin{figure}[tb]
\vspace{-0.08cm}
\mbox{\hspace{-0.4cm}\includegraphics[width=0.52\textwidth]{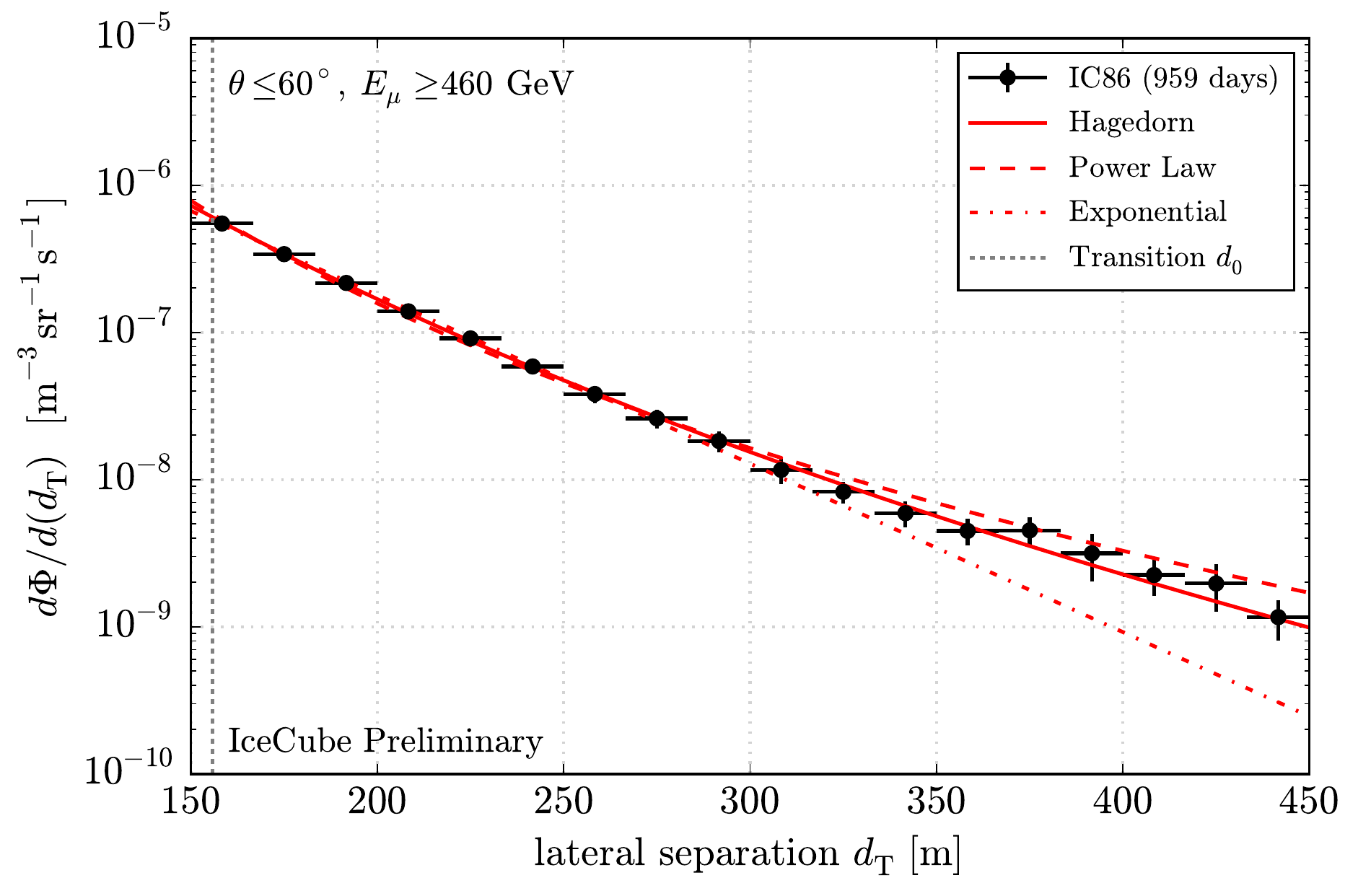}}
\vspace{-0.8cm}
\caption{Lateral separation distribution of muons with energy above $460\,\mathrm{GeV}$ and zenith angle $\theta\leq 60^\circ$, obtained from three years of IceCube data. Also shown is the corresponding Hagedorn fit of the form of Equation (\ref{eq:hagedorn}), as well as an exponential and a power law fit for comparison.}
\label{fig:LS_dT1}
\vspace{-0.3 cm}
\end{figure}

\begin{figure*}[tb]
\centering
\mbox{\hspace{-0.3cm}\includegraphics[width=0.85\textwidth]{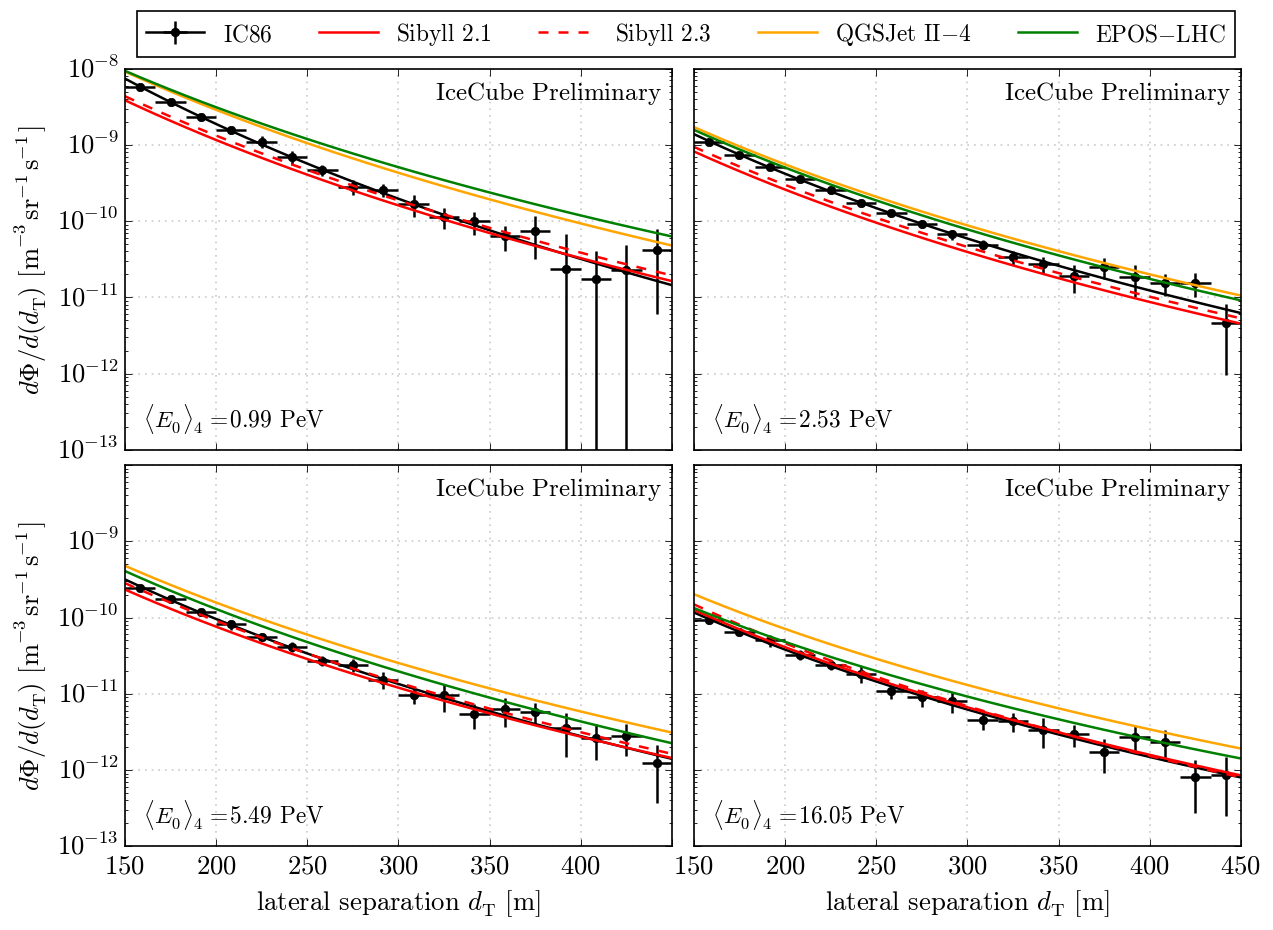}}
\vspace{-0.2cm}
\caption{Lateral separation distributions  of muons with energy above $460\,\mathrm{GeV}$ and zenith angle $\theta\leq 60^\circ$ for different primary energy bins. The corresponding mean energies are given in the figures. Also shown are Hagedorn fits of the form of Equation (\ref{eq:hagedorn}) (black lines) and CORSIKA predictions, using different hadronic models and a H3a primary spectrum assumption.}
\label{fig:LS_dT2}
\vspace{-0.3 cm}
\end{figure*}

The bright muon bundle together with the isolated LS muon form a distinct double-track signature in IceCube. The lateral distance of the LS muon to the shower core is approximately given by
\begin{linenomath}
\begin{align}
d_\mathrm{T}\simeq \frac{p_\mathrm{T}\cdot H}{E_\mu\cdot \cos(\theta)}\, ,
\label{eq:dT}
\end{align}
\end{linenomath}
where $p_\mathrm{T}$ is the transverse momentum of the muon, $E_\mu$ is the muon energy, $\theta$ is the zenith angle direction, and $H$ is the altitude of hadron production. Figure \ref{fig:LS_dT1} shows the lateral separation distribution of muons (after background subtraction) on surface level, obtained from three years of IceCube data. In order to derive the distributions at surface level, effective areas obtained from CORSIKA simulations are used. The event selection is based on a previous analysis, which used one year of data from IceCube in its 59-string configuration and is described in Ref. \cite{IceCube_LS1}. Also shown in Figure \ref{fig:LS_dT1} is a QCD-inspired Hagedorn fit \cite{Hagedorn} of the form
\begin{linenomath}
\begin{align}
\frac{d\Phi}{d(d_\mathrm{T})}= \alpha\cdot \left( 1+\frac{d_\mathrm{T}}{d_0} \right)^{-\beta} ,
\label{eq:hagedorn}
\end{align}
\end{linenomath}
with $\alpha$, $\beta$, and $d_0$ being free parameters. The resulting best fit parameters can be found in Table \ref{tab:hagedorn}. This functional form behaves like an exponential for $d_\mathrm{T}/d_0\rightarrow 0$ and describes a power law for $d_\mathrm{T}/d_0\rightarrow \infty$, with the transition around $d_0$. Also shown are fits assuming a pure exponential and a simple power law. The Hagedorn function describes the experimental distribution well ($\chi^2/\mathrm{ndof}=20.16/16$), with the transition from soft to hard interactions at around $d_0=(157.3\pm 43.0)\,\mathrm{m}$. In contrast, the pure exponential and power law fits are in poor agreement with the data, especially towards large separations ($\chi^2/\mathrm{ndof}=97.92/16$ and $\chi^2/\mathrm{ndof}=53.60/16$ respectively). Within uncertainties, the measured distribution, as well as the resulting fit parameters, are in agreement with previous results \cite{IceCube_LS1}.

\begin{table}[b]
\begin{tabular*}{0.48\textwidth}{c | ccc}
\hline
$\langle E_0\rangle$ & $\alpha$ & $\beta$ & $d_0$  \\\hline
$4.08\,\mathrm{PeV}$ & $236.3\pm 145.3$ & $9.7\pm 1.1$ & $157.3\pm 43.0$  \\\hline
$0.99\,\mathrm{PeV}$ & $1.55\pm 1.23$ & $10.2\pm 1.9$ & $186.6\pm 77.4$ \\
$2.53\,\mathrm{PeV}$ & $0.12\pm 0.10$ & $8.1\pm 1.7$ & $167.5\pm 81.6$ \\
$5.49\,\mathrm{PeV}$ & $0.03\pm 0.03$ & $8.1\pm 1.9$ & $170.9\pm 93.5$ \\
$16.05\,\mathrm{PeV}$ & $0.01\pm 0.01$ & $6.9\pm 1.8$ & $133.8\pm 97.4$ \\\hline
\end{tabular*}
\caption{Hagedorn fit parameters $\alpha$ (in $10^{-6}\,\mathrm{m}^{-3}\mathrm{sr}^{-1}\mathrm{s}^{-1}$), $\beta$, and $d_0$ (in m), as defined in Equation (\ref{eq:hagedorn}). Corresponding to the fits shown in Figure \ref{fig:LS_dT1} ($4.08\,\mathrm{PeV}$) and Figure \ref{fig:LS_dT2}.}
\label{tab:hagedorn}
\end{table}

Using an energy estimator based on the truncated mean of the energy losses along the reconstructed bundle track \cite{IceCube_trunc}, the primary energy of the cosmic ray air shower is derived for each event. The resulting mean cosmic ray energy of events shown in Figure \ref{fig:LS_dT1} is approximately $\langle E_0 \rangle=4.08\,\mathrm{PeV}$. The lateral separation distributions for four primary energy bins are shown in Figure \ref{fig:LS_dT2}, with the corresponding Hagedorn fits of the form of Equation (\ref{eq:hagedorn}) shown as black lines. The resulting best fit parameters are given in Table \ref{tab:hagedorn}. Also shown are predictions from different hadronic interaction models, Sibyll 2.1 \cite{Sibyll21}, Sibyll 2.3 \cite{Sibyll23}, QGSJet II-4 \cite{QGSJet}, and EPOS-LHC \cite{EPOS}, obtained from CORSIKA simulations at surface level, using an H3a primary flux assumption. While QGSJet II-4 and EPOS-LHC predict larger LS muon fluxes with flatter lateral separation distributions, Sibyll models are in good agreement with experimental data, especially towards higher primary energies.

\section{Angular distributions}
\label{Sec:4}

\begin{figure}[!t]
\vspace{-0.3 cm}
\mbox{\hspace{-0.2cm}\includegraphics[width=0.5\textwidth]{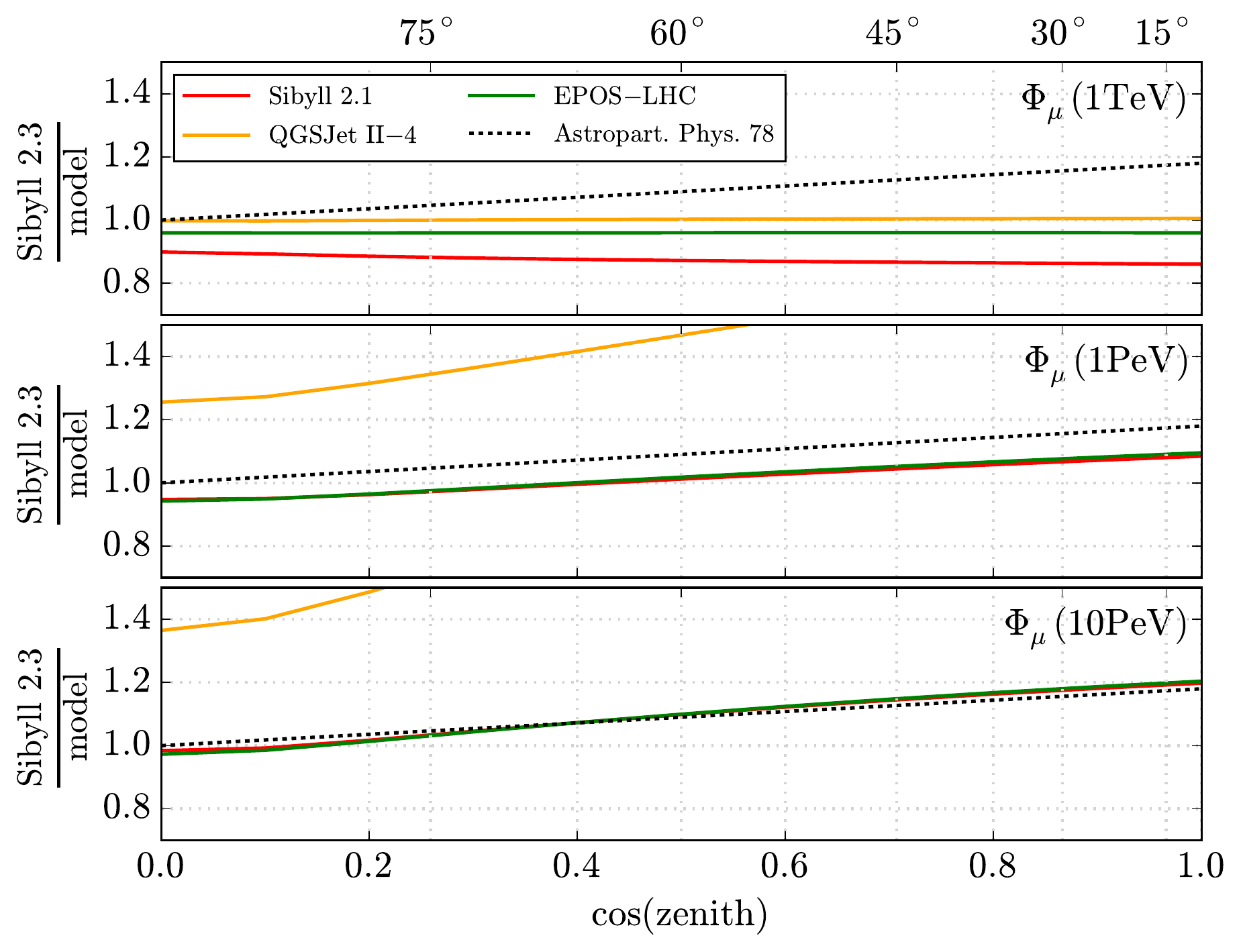}}
\vspace{-0.8 cm}
\caption{Ratio of the atmospheric muon flux between Sibyll 2.3 predictions and various other models for a given muon energy, as a function of the zenith angle. The muon fluxes are generated using MCEq \cite{MCEq}. Also shown for comparison is the zenith angle discrepancy $f_\mathrm{HE}(\theta)$, as reported in Ref. \cite{IceCube_HE1} and defined in the text.}
\label{fig:MCEq_zen}
\vspace{-0.3 cm}
\end{figure}

\begin{figure}[!t]
\vspace{-0.3 cm}
\mbox{\hspace{-0.2cm}\includegraphics[width=0.49\textwidth]{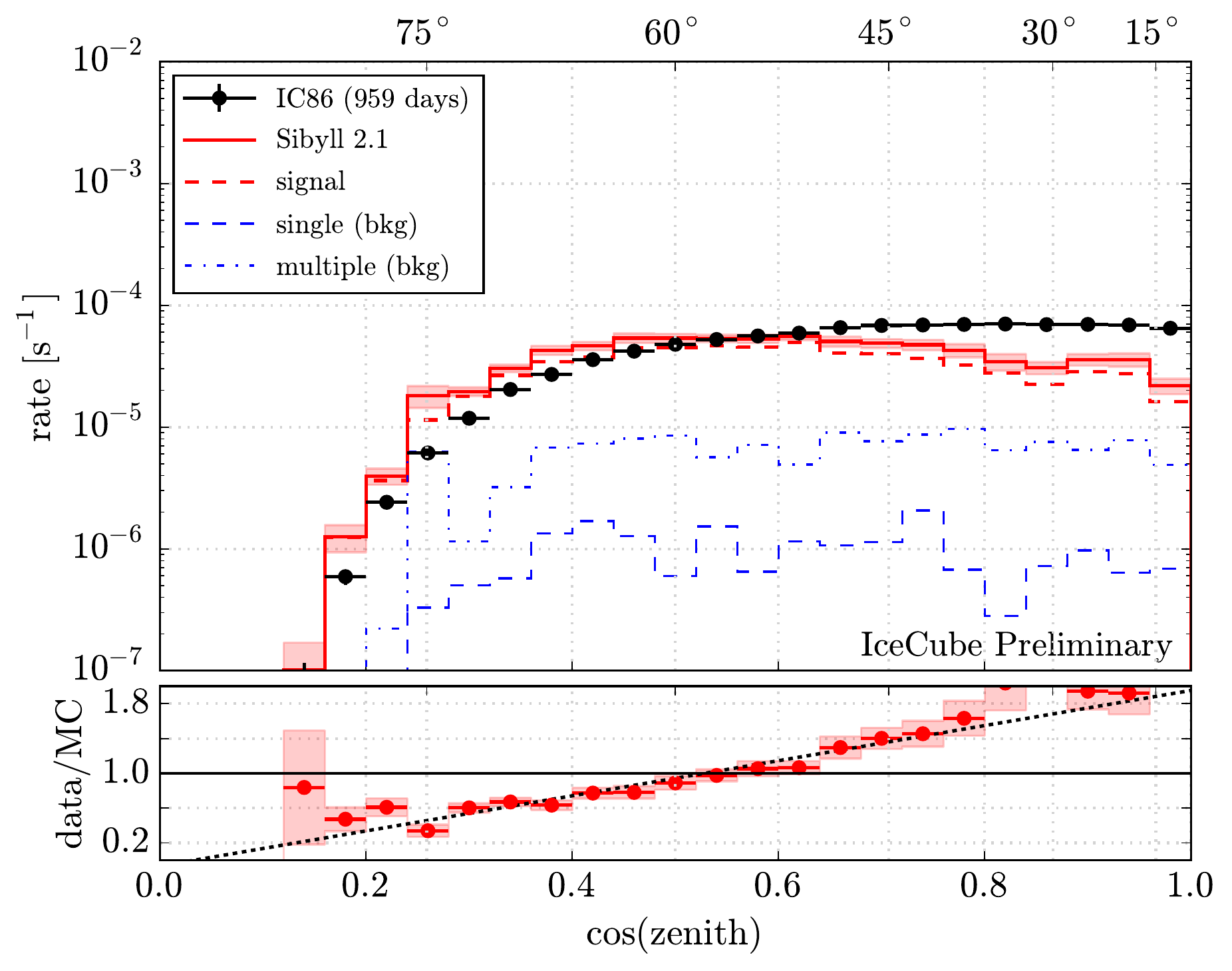}}
\vspace{-0.85 cm}
\caption{Zenith angle distribution of LS muon events at final analysis level in the ice for experimental data and CORSIKA simulations with Sibyll 2.1. Blue lines indicate background contributions from mis-reconstructed bundle events (single) and in-time coincident showers (multiple). The bottom panel shows the ratio between experimental data and simulations.}
\label{fig:LS_zen}
\vspace{-0.9 cm}
\end{figure}

As reported in Ref. \cite{IceCube_HE1}, a discrepancy in the angular distribution of high-energy muons in the ice between experimental data and CORSIKA simulations using Sibyll 2.1 is observed. While negligible at trigger level the disagreement becomes significant at final analysis level where it can be parameterized as $f_\mathrm{HE}(\theta)=1.0+0.18\cdot \cos(\theta)$, as described in Ref. \cite{IceCube_HE1}. Dedicated studies of the ice properties, the efficiency and angular acceptance of the optical modules, various cosmic ray flux assumptions, and other effects do not provide an explanation for the observed angular mismatch. Figure \ref{fig:MCEq_zen} shows the ratio of predictions obtained from Sibyll 2.3 and various other models. The flux predictions are generated using MCEq \cite{MCEq} and shown for different muon energies. For comparison, the discrepancy observed in Ref. \cite{IceCube_HE1} is also shown (dotted line). Towards high energies the previously observed discrepancy shows qualitative similarities with the differences between Sibyll 2.3 and other hadronic models. Although this simplified picture does not allow final conclusions, it provides evidence that the observed discrepancies may be related to hadronic modeling. Figure \ref{fig:LS_zen} shows the zenith angle distribution of LS muons in the ice at final analysis level with a significant diagreement between experimental data and CORSIKA simulations. The data to Monte Carlo ratio can be parametrized as $f_\mathrm{LS}(\theta)=-0.06+2.02\cdot \cos(\theta)$. This angular mismatch was previously reported in Ref. \cite{IceCube_LS1} and can not be explained by model uncertainties. This is because LS muons typically have energies in the $\mathrm{TeV}$ range where the differences between Sibyll 2.3 predictions and other models are negligible (see top Figure \ref{fig:MCEq_zen}). Various further studies did not find any explanation for the observed mismatch and the disagreement is not yet understood.

\section{Conclusions}
\label{Sec:5}

The energy spectrum of high-energy muons measured between 10 TeV and 1 PeV in IceCube has been presented. A simple power law fit has a spectral index of $-3.76\pm0.02$ and the best prompt estimate is $4.75\times \Phi_\mathrm{ERS}$, assuming an H3a primary spectrum. However, the systematic uncertainties of this measurement are significant and an improved analysis is in preparation. 
In addition, the lateral separation distributions of TeV muons between $150\,\mathrm{m}$ and $450\,\mathrm{m}$ has been presented for several cosmic ray energies. The resulting distributions are in good agreement with previous measurements and CORSIKA simulations. The expected transition from an exponential to a power law behavior is observed at around $d_0=(157\pm43)\,\mathrm{m}$. 
The corresponding angular distributions of atmospheric muons show a mismatch between experimental data and CORSIKA simulations. Although Sibyll 2.3 provides evidence

\newpage

\noindent   that the discrepancy at the highest energies may be related to the interaction model, this disagreement is not yet understood and thus further studies using recent hadronic models are in progress.

\end{document}